# Radii of the $E_8$ Gosset Circles as the Mass Excitations in the Ising Model

**Mehmet Koca**[a)] **and Nazife Ozdes Koca**[b)]

Department of Physics, College of Science, Sultan Qaboos University
P. O. Box 36, Al-Khoud 123, Muscat, Sultanate of Oman

## ABSTRACT

The Zamolodchikov's conjecture implying the exceptional Lie group $E_8$ seems to be validated by an experiment on the quantum phase transitions of the 1D Ising model carried out by the Coldea et. al. The $E_8$ model which follows from the affine $E_8$ Toda field theory predicts 8 bound states with the mass relations in the increasing order

$$m_1, m_2 = \tau m_1, m_3, m_4, m_5, m_6 = \tau m_3, m_7 = \tau m_4, m_8 = \tau m_5, \text{ where } \tau = \frac{1+\sqrt{5}}{2}$$ represents the golden ratio. Above relations follow from the fact that the Coxeter group $W(H_4)$ is a maximal subgroup of the Coxeter-Weyl group $W(E_8)$. These masses turn out to be proportional to the radii of the Gosset's circles on the Coxeter plane obtained by an orthogonal projection of the root system of $E_8$. We also note that the masses $m_1, m_3, m_4, m_5$ correspond to the radii of the circles obtained by projecting the vertices of the 600-cell, a 4D polytope of the non-crystallographic Coxeter group $W(H_4)$.

A special non-orthogonal projection of the simple roots on the Coxeter plane leads to exactly the numerical values of the masses of the bound states as 0.4745, 0.7678, 0.9438, 1.141, 1.403, 1.527, 1.846, and 2.270. We note the striking equality of the first two numerical values to the first two masses of the bound states determined by the Coldea et. al.

Keywords: Zamolodchikov conjecture, Ising model, Coxeter-Weyl group $W(E_8)$, Coxeter plane, Coxeter graph $I_2(30)$, Gosset circles

[a)] electronic-mail: kocam@squ.edu.om
[b)] electronic-mail: nazife@squ.edu.om

The experimental observation performed by Radu Coldea et.al.[1] on the quantum critical phase transition of the 1D Ising model represented by the crystal $CoNb_2O_6$ (cobalt niobate) created excitement for it could validate the Zamolodchikov's conjecture [2] related to the $E_8$ Lie algebra. Using the S-matrix approach Zamolodchikov predicted that the masses of the bound states of the 1D Ising model can be determined, in the increasing order, as

$$m_1, m_2 = \tau m_1, m_3, m_4, m_5, m_6 = \tau m_3, m_7 = \tau m_4, m_8 = \tau m_5 \tag{1}$$

where $m_1$ is a free parameter and $m_3, m_4, m_5$ and $\tau$ are given by

$$m_3 = 2m_1 \cos\frac{\pi}{30}, m_4 = 2m_2 \cos\frac{7\pi}{30}, m_5 = 2m_2 \cos\frac{2\pi}{15} \text{ and } \tau = 2\cos\frac{\pi}{5} = \frac{1+\sqrt{5}}{2}. \tag{2}$$

The team led by R. Coldea have indications for the first two mass excitations with the masses $m_1 \approx 0.5$ meV and $m_2 \approx \tau m_1$.

The integrable field theories and the 1+1 dimensional conformal field theories were very attractive research areas during the late 80's and early 90's. It was shown that, in the continuum limit, the 1D Ising model perturbed with an external magnetic field, can be represented by the affine Toda field theory of $E_8$ [3] which corresponds to the perturbed Toda field theory of $E_8$. The Lagrangian of the affine Toda field theory is given by

$$L = \frac{1}{2}\partial^\mu \phi^a \partial_\mu \phi^a - \frac{m^2}{\beta}\sum_{i=0}^{l} n_i \exp\{\beta\alpha_i.\phi\} \tag{3}$$

where the eigenvalues of the mass matrix $M^{ab} = m^2 \sum_{i=0}^{l} n_i \alpha_i^a \alpha_i^b$ can be related to the matrix $N_{ij} = m^2 \sum_{a=1}^{8} n_i \alpha_i^a \alpha_j^a$ [4]. Here $\alpha_i$ $(i=1,2,...,8)$ are the simple roots and $n_i$ denote the coefficients of the highest root of the $E_8$ Lie algebra [5]. One should also recall that the coefficients $n_i$, together with 1, correspond to dimensions of the irreducible representations of the binary icosahedral group [6]. If the characteristic polynomial of the mass matrix is given by the function $f(x)$ the characteristic polynomial of the singular matrix $N$ is given by $xf(x)$. Square roots of the eigenvalues of the matrix $N$, besides zero eigenvalue, lead to the masses given in equations (1-2). P. Dorey [7] has shown that the radii of the circles obtained by projecting the root systems of the simple laced Lie Algebras are proportional to the masses of the scalar particles of the Affine Toda field theories. A similar work was carried out earlier by H.W. Braden [8] in the context of "fusing rule". B. Kostant [9] has shown that the mass matrix can be associated with an operator defined in the dual bases and its eigenvalues correspond to the masses of the scalar particles obtained from the Affine Toda field theory. In particular, he has shown that the radii of 8 Gosset circles obtained by projecting on the Coxeter plane 240 vertices of the Gosset polytope are proportional to the masses given in (1).

In what follows we derive the same results with a simpler technique by projecting the $E_8$ lattice to the Coxeter plane defined by the simple roots of its maximal dihedral subgroup $D_{30}$ corresponding to the Coxeter diagram $I_2(30)$. We also note that the four of the masses related to the other four masses by a factor of $\tau$ happen to be that the Coxeter-Weyl group $W(E_8)$ admits the non-crystallographic Coxeter group $W(H_4)$ as a maximal subgroup.

Every Coxeter-Weyl group has a maximal dihedral group, up to conjugation, with an order $2h$ where $h$ is the Coxeter number. The dihedral group of order $2h$ can be represented by the Coxeter graph $I_2(h)$. Denote by $\alpha_i$ and $r_i$ $(i=1,2,...,8)$ the simple roots and the reflection generators of $W(E_8)$ respectively as shown in Fig.1.

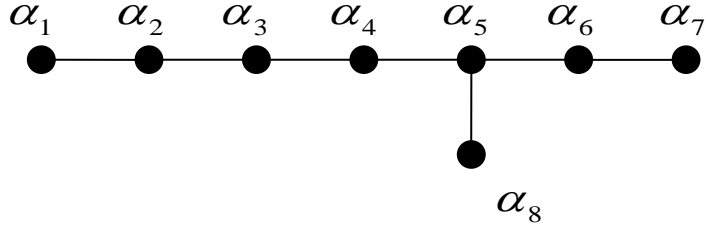

Fig.1. Coxeter-Dynkin diagram of $E_8$.

Similarly, let us denote the simple roots and the reflection generators of the $H_4$ respectively by $\beta_i$ and $R_i$ or $\beta_i'$ and $R_i'$ ($i=1, 2, 3, 4$) as shown in Fig.2.

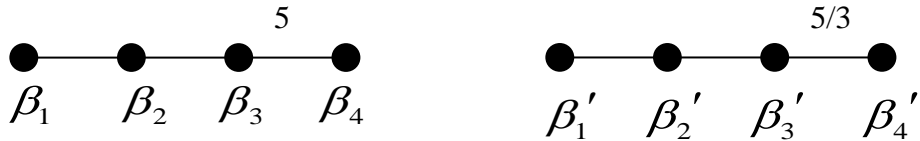

Fig.2. Two Coxeter diagrams of $H_4$.

Note that the angle between the third and the fourth simple roots of $H_4$ is $144^0$. One can write the respective generators and the simple roots of the Coxeter group $W(H_4)$ in terms of the reflection generators and the simple roots of $W(E_8)$ as follows [10]

$$R_1 = r_1 r_7, R_2 = r_2 r_6, R_3 = r_3 r_5, R_4 = r_4 r_8, \ \beta_a = x_{ai}\alpha_i, \ (a=1,2,3,4), (i=1,2,...,8). \quad (4)$$

Acting the generators on the simple roots $\beta_a$ in (4) we determine the coefficients

$$\beta_1 = \frac{1}{\sqrt{2+\tau}}(\alpha_1 + \tau\alpha_7), \beta_2 = \frac{1}{\sqrt{2+\tau}}(\alpha_2 + \tau\alpha_6),$$
$$\beta_3 = \frac{1}{\sqrt{2+\tau}}(\alpha_3 + \tau\alpha_5), \beta_4 = \frac{1}{\sqrt{2+\tau}}(\tau\alpha_4 + \alpha_8). \quad (5)$$

Had we chosen the angles between the third and the fourth roots as $72^0$ then the simple roots would read

$$\beta'_1 = \frac{1}{\sqrt{2+\sigma}}(\alpha_1 + \sigma\alpha_7), \beta'_2 = \frac{1}{\sqrt{2+\sigma}}(\alpha_2 + \sigma\alpha_6),$$
$$\beta'_3 = \frac{1}{\sqrt{2+\sigma}}(\alpha_3 + \sigma\alpha_5), \beta'_4 = \frac{1}{\sqrt{2+\sigma}}(\sigma\alpha_4 + \alpha_8) \quad (6)$$

Note that all the roots are normalized to $\sqrt{2}$ and $\sigma = \frac{1-\sqrt{5}}{2}$. The generators corresponding to the simple roots of (6) also generate the Coxeter group $W(H_4)$.

These two sets of root systems constitute two orthogonal systems in the 8-dimensional Euclidean space. Writing two sets of equations in the matrix form

$$\beta_i = g_{ij}\alpha_j \ (i,j = 1,2,...,8) \text{ where } \beta_5 = \beta'_1, \beta_6 = \beta'_2, \beta_7 = \beta'_3, \beta_8 = \beta'_4$$

one can check that the Cartan matrix of $E_8$ is block diagonalized where the upper and the lower block matrices represent the Cartan matrices of the Coxeter diagram $H_4$ with the angles between the third and the fourth roots of the upper and lower blocks are respectively $144^0$ and $72^0$[11].

Let us now define reflection generators and simple roots of the Coxeter diagram $I_2(30)$ by

$$S_1 = R_1 R_3, \ S_2 = R_2 R_4, \ \gamma_i = z_{ia}\beta_a, \ (i = 1,2; \ a = 1,2,3,4), \quad (7)$$

with the relations $(\gamma_1, \gamma_1) = (\gamma_2, \gamma_2) = 2, (\gamma_1, \gamma_2) = -c, c = 2\cos\frac{\pi}{30}$.

Using the fact that $S_1\gamma_1 = -\gamma_1$ and $S_2\gamma_2 = -\gamma_2$, the expression in (7) can further be reduced to the equations

$$\gamma_1 = z_1\beta_1 + z_3\beta_3, \ \gamma_2 = z_2\beta_2 + z_4\beta_4. \quad (8)$$

Now using the relations $S_1\gamma_2 = \gamma_2 + c\gamma_1, S_2\gamma_1 = \gamma_1 + c\gamma_2$ one can determine the coefficients as

$$z_2 = cz_1; \; z_3 = (c^2-1)z_1 = (2\tau\cos\frac{2\pi}{15})z_1; \; z_4 = \frac{\tau(c^2-1)}{c}z_1 = (2\tau\cos\frac{7\pi}{30})z_1. \tag{9}$$

Their numerical values can be determined through the root normalizations as

$z_1 \approx 0.3204, \; z_2 = 0.6373, \; z_3 = 0.9473, \; z_4 = 0.7706.$

Note the relations in (9) can be obtained from the eigenvalue equation

$$\begin{pmatrix} 0 & 1 & 0 & 0 \\ 1 & 0 & 1 & 0 \\ 0 & 1 & 0 & \tau \\ 0 & 0 & \tau & 0 \end{pmatrix} \begin{pmatrix} z_1 \\ z_2 \\ z_3 \\ z_4 \end{pmatrix} = c \begin{pmatrix} z_1 \\ z_2 \\ z_3 \\ z_4 \end{pmatrix}, \tag{10}$$

where the matrix in (10) is $(2I - C_{H_4})$ with the matrix $C_{H_4}$ is the Cartan matrix of the $H_4$ diagram. Due to the Perron-Frobenius theorem the components of the eigenvector are positive since $c$ is the largest eigenvalue of the matrix in (10). Now the expression in (8) can be rewritten in terms of the simple roots of $E_8$ as follows

$$\gamma_1 = \frac{1}{\sqrt{2+\tau}}[z_1(\alpha_1 + \tau\alpha_7) + z_3(\alpha_3 + \tau\alpha_5)], \; \gamma_2 = \frac{1}{\sqrt{2+\tau}}[z_2(\alpha_2 + \tau\alpha_6) + z_4(\tau\alpha_4 + \alpha_8)]. \tag{11}$$

We note that these 8 coefficients $z_1, \tau z_1, z_2, \tau z_2, z_3, \tau z_3, z_4, \tau z_4$ can be directly obtained as the components of the eigenvector of the matrix $(2I - C_{E_8})$ corresponding to the largest eigenvalue $c = 2\cos(\frac{\pi}{30})$. The plane defined by the simple roots $\gamma_1$ and $\gamma_2$ of the Coxeter graph is known as the Coxeter plane of the $E_8$ lattice. Any vector of the $E_8$ lattice can be projected on to the plane using the simple roots in (11).

We will discuss two cases of projections: orthogonal projections and non-orthogonal projections. For an orthogonal projection one can define the orthogonal unit vectors in the Coxeter plane as follows

$$\hat{i} = \frac{\gamma_1}{\sqrt{2}}, \; \hat{j} = \frac{1}{\sin\frac{\pi}{30}}(\frac{\gamma_2}{\sqrt{2}} + \frac{c}{2}\frac{\gamma_1}{\sqrt{2}}). \tag{12}$$

The 240 vertices of the $E_8$ Gosset polytope are mapped on the Coxeter plane by using the orthogonal projection. One can prove that $E_8$ root system can be decomposed under the dihedral subgroup $D_{30} = \langle S_1, S_2 \rangle$ into 8 orbits each containing 30 roots and each orbit is characterized by one simple root $\alpha_i$ ($i = 1,2,...,8$). Under the orthogonal projection each $\alpha_i$-orbit determines a circle of radius $r_i$ given in the increasing order by

$$r_1 = |\alpha'_1| = \sqrt{\frac{2}{2+\tau}} z_1,\ r_2 = |\alpha'_7| = \tau r_1,\ r_3 = |\alpha'_2| = \sqrt{\frac{2}{2+\tau}} z_2,\ r_4 = |\alpha'_8| = \sqrt{\frac{2}{2+\tau}} z_4,$$

$$r_5 = |\alpha'_3| = \sqrt{\frac{2}{2+\tau}} z_3,\ r_6 = |\alpha'_6| = \tau r_3,\ r_7 = |\alpha'_4| = \tau r_4,\ r_8 = |\alpha'_5| = \tau r_5 \quad (13)$$

where the norm of the image of the simple root $\alpha_i$ on the Coxeter plane is denoted by $|\alpha'_i|$. Ratios of these radii satisfy the same ratio satisfied by the masses of the bound states. We also note that the norms of images of the fundamental weight vectors satisfy the same ratios of the masses. Therefore one can associate the radii of the Gosset circles up to a scale factor to the masses of the quasiparticle states of the Ising model. The relations in (13) also imply that the components of the eigenvector of the Cartan matrix of $E_8$ corresponding to the highest eigenvalue are proportional to the masses of the exited states.

The image of the $E_8$ lattice can also be determined by a non-orthogonal projection. In this case vectors of a given $\alpha_i$-orbit generated by the maximal dihedral subgroup $D_{30} = \langle S_1, S_2 \rangle$ do not map on the same circle. Nevertheless the images of the $E_8$ lattice vectors form concentric circles on the Coxeter plane with increasing radii. A particular projection of the $E_8$ lattice on the Coxeter plane can be made by computing the components of the lattice vectors on the root vectors $\gamma_1$ and $\gamma_2$ of the dihedral group $D_{30} = \langle S_1, S_2 \rangle$. The image of an arbitrary $E_8$ lattice vector on the Coxeter plane can be written as

$\Lambda' = \Lambda_1 \frac{\gamma_1}{\sqrt{2}} + \Lambda_2 \frac{\gamma_2}{\sqrt{2}}$ with the norm of the image $|\Lambda'| = (\Lambda_1^2 + \Lambda_2^2 - \Lambda_1 \Lambda_2 c)^{1/2}$.

With this projection the images of the simple roots will lie on different circles with radii in the increasing order

$$r_1 = |\alpha'_1| = \sqrt{\frac{4+3c^2}{2(2+\tau)}} z_1,\ r_2 = |\alpha'_7| = \tau r_1,\ r_3 = |\alpha'_2| = \sqrt{\frac{4+3c^2}{2(2+\tau)}} z_2,\ r_4 = |\alpha'_8| = \sqrt{\frac{4+3c^2}{2(2+\tau)}} z_4,$$

$$r_5 = |\alpha'_3| = \sqrt{\frac{4+3c^2}{2(2+\tau)}} z_3,\ r_6 = |\alpha'_6| = \tau r_3,\ r_7 = |\alpha'_4| = \tau r_4,\ r_8 = |\alpha'_5| = \tau r_5. \quad (14)$$

Numerical values of these radii are 0.4745, 0.7678, 0.9438, 1.141, 1.403, 1.527, 1.846 and 2.270. It seems that these radii directly determine the masses of the bound states!

The first and the second radii are in agreement with the first two masses of the bound states determined by the Coldea et.al. We have no idea whether this is just a numerical coincidence or there is a deeper relation between the masses of these states and the radii of the circles just determined by the above non-orthogonal projection of the simple roots. We hope that the above simple algebraic analysis will shed some light on the evidence of the $E_8$ symmetry in the quantum phase transition of the 1D Ising model. One should also note that the Coxeter plane of an arbitrary Weyl-Coxeter group can be determined in a similar manner.

We thank Ramazan Koç for discussions.